\documentclass[aps,prb,floatfix,preprint]{revtex4}
\usepackage{amssymb}
\usepackage{amsmath}
\usepackage{graphicx}

\begin{document}

\title{\Huge Photon mediated interaction between distant quantum dot circuits}

\author{M.R. Delbecq, L.E. Bruhat, J.J. Viennot, S. Datta, A. Cottet and T. Kontos
\footnote {To whom correspondence should be addressed :
kontos@lpa.ens.fr}}
\affiliation{\normalsize{Laboratoire Pierre Aigrain, Ecole Normale Sup\'erieure, CNRS UMR 8551, Laboratoire associ\'e aux universit\'es Pierre et Marie Curie et Denis Diderot, 24, rue Lhomond, 75231 Paris Cedex 05,
France}\\}

\date{\today}

\begin{abstract}

Engineering the interaction between light and matter is an important goal in the emerging field of quantum opto-electronics. Thanks to the use of cavity quantum electrodynamics architectures, one can envision a fully hybrid multiplexing of quantum conductors. Here, we use such an architecture to couple two quantum dot circuits . Our quantum dots are separated by 200 times their own size, with no direct tunnel and electrostatic couplings between them. We demonstrate their interaction, mediated by the cavity photons. This could be used to scale up quantum bit architectures based on quantum dot circuits or simulate on-chip phonon-mediated interactions between strongly correlated electrons.

\end{abstract}

\maketitle

Cavity QED allows one to study the interaction between light and matter at the most elementary level, for instance,  by using Rydberg atoms
coupled to cavity photons \cite{Raimond:01}. Recently, it has become possible to perform similar experiments on-chip, by using artificial two-level
systems made from superconducting circuits instead of atoms \cite{Wallraff:04}. This circuit-QED offers unexplored potentialities, since other
degrees of freedom than those of superconducting circuits could be used \cite{Schoelkopf:08,Trif:08,Cottet:10,Jin:11,Cottet:12,Kubo:10,Schuster:10,Bergenfeld:12}, and in particular, those of quantum dots \cite{Delbecq:11,Frey:12,Petta:12,Toida:12}. Such a hybrid circuit QED would allow one to study a large variety of situations
not accessible with standard cavity QED, owing to the versatility of nanofabricated circuits \cite{Cottet:12,Trif:12}.

A highly desired functionality required for the use of quantum dots for quantum information processing is the controlled coupling between distant quantum dots. This would allow one to scale up the existing spin quantum bit architectures\cite{Petta:05,Nowack:07,Foletti:09}. Among the possible methods for coupling quantum dots \cite{Petta:05,Hermelin:11,McNeil:11}, the use of photons is particularly attractive because they can mediate a coherent interaction over macroscopic distances \cite{Raimond:01,Majer:07,Burkard:06}. Recently,  the coupling of \textit{single} quantum dot circuits to microwave photons has been demonstrated. So far, this architecture has been used to directly measure the electron-photon coupling strength \cite{Delbecq:11,Frey:12,Toida:12} or to read-out capacitively a spin quantum bit \cite{Petta:12}.

Here, we demonstrate a distant interaction by embedding two quantum dot circuits in a microwave cavity. Our quantum dots are separated by 200 times their own size, with no direct tunnel and/or electrostatic coupling. We use one of the two dots as a detector or perform a joint read-out of the two quantum dots with the microwave cavity in order to probe the distant interaction. The spectroscopy of the detector dot is shown to display a dispersion as well as crossings of charge states as a function of its own gate and that of the \textit{other distant dot}. The joint read-out allows us to probe in a complementary fashion the distant coupling scheme through the phase of the microwave signal.

\textbf{Results}

\textbf{Sample geometry.}
The device presented here is shown in figure 1a and 1c (see Methods). The two quantum dots, made using \textit{different} and \textit{well separated} single wall carbon nanotubes, are first studied using the traditional transport spectroscopy. The colorscale plots of the differential conductance of dot 1 (QD1) and dot 2 (QD2) as a function of their gate and source-drain bias are represented in figure 1d and 1e respectively. For QD1, the conductance is close to $2\textrm{e}^2/\textrm{h}$ and displays modulations with the characteristic checker-board pattern of an electronic Fabry-Perot interferometer \cite{Liang:02}. The vertical distance between the bright spots allows us to determine a level spacing of about $7\textrm{meV}$. This corresponds to a length of about $200\textrm{nm}$, which is slightly smaller than the lithographically defined length. The quantum dot 1 can be considered as an open quantum dot. The differential conductance of QD2 is much smaller and peaks only up to $0.03 \times 2\textrm{e}^2/\textrm{h}$. The colorscale plot of its differential conductance displays Coulomb diamonds and excited states characteristic of the Coulomb blockade regime \cite{Cobden:02}. In particular, one can read off a charging energy of about $12 \textrm{meV}$. Another important parameter for both dots is the width $\Gamma_{1(2)}$ of their energy levels. From the colorscale plots, one finds, $\Gamma_1 \approx 7 \textrm{meV}$ and $\Gamma_2 \approx 1 \textrm{meV}$. This shows that the two electronic systems still relax much faster than the characteristic period of the photonic mode, which corresponds to $5.75 \textrm{GHz} \cong 30 \mu \textrm{eV}$ as shown in figure 1b.

%The electronic systems are both capacitively coupled to the cavity photons \cite{Delbecq:11,Frey:12,Toida:12}. In a similar fashion as in cavity QED, %their effective spectrum can be tuned by changing the average number of photons stored in the cavity \cite{Raimond:01}, or equivalently here, by the %amplitude of the cavity electric field. The latter is controlled through the microwave power sent to the input port of the cavity. The electronic %systems can also be affected by virtual photonic processes, in the absence of any microwave drive. The latter processes will be crucial for %understanding the photon mediated interaction between the two quantum dots.
\textbf{Calibration of electron-photon coupling for each dot.}
Before going to the main result of the present work, which is the distant interaction between the above described quantum dots, it is essential to calibrate the electron-photon coupling for each dot. In order to do this, we apply a strong, classical, microwave drive to the cavity at the resonance frequency. We first focus on the closed dot which allows us to obtain the highest spectral resolution since it has the smallest coupling rate to its reservoirs. The colorscale plots of the differential conductance of QD2 as a function of $V_\textrm{g2}$ and $V_\textrm{sd2}$ for no microwave power, $-45 \textrm{dBm}$, and $-35 \textrm{dBm}$ at the input of the cavity are presented in figure 2a, 2b and 2c respectively. The microwave field splits dynamically the charge degeneracy line crossing at about $V_\textrm{g2}=-11.7\textrm{V}$. This splitting is about $3\textrm{meV}$ at $-45\textrm{dBm}$ and increases up to $10\textrm{meV}$ at $-35 \textrm{dBm}$.  One can understand classically the observed phenomenon by using the fact that the dot's electrons have a relaxation rate to the electrodes which is about 45 times higher than the microwave frequency of the cavity. The AC field of the resonator modulates the dot's energy levels with an amplitude $\lambda V_\textrm{AC}$ and frequency $f_0$. The averaging of this modulation leads to split conductance peaks (dashed lines) at $E_0 \pm \lambda V_\textrm{AC}$, as depicted schematically in figure 2d. The differential conductance $G_\textrm{on}(V_\textrm{g},V_\textrm{sd})$ at finite microwave drive can be obtained by assuming that the microwave field only couples to the electronic levels, leading to : $G_\textrm{on}(V_\textrm{g},V_\textrm{sd})=1/2 \pi \int_{0}^{2 \pi} d \theta G_\textrm{off}(V_\textrm{g}+\lambda V_\textrm{AC} \cos \theta,V_\textrm{sd})$, $\lambda$ characterizing the electron-photon coupling strength, $V_\textrm{AC}$ being the amplitude of the AC voltage at the resonator frequency $f_0$ and $G_\textrm{off}(V_\textrm{g},V_\textrm{sd})$ being the differential conductance of QD2 with no microwave power. The results of this procedure is shown in figure 2e and 2f respectively for $\lambda=15.5$, using the \textit{measured} data of figure 2a. The conductance peak splitting and in particular the peculiar diamond shaped region emerging at the charge degeneracy point around $V_\textrm{g2}=-11.7\textrm{V}$ are very well accounted for. Note that not all the features in figure 2b and 2c are reproduced with our procedure, indicating that some of the excited states might have a different coupling to the photons than the ground state \cite{Palyi:11}.

The specific origin of the conductance peak splitting can be confirmed by studying its dependence as a function of the power of the microwave signal at the input of the cavity, or equivalently the average number of photons stored in the cavity. Although we have used a classical field so far, such a study allows us to determine accurately the electron-photon coupling terms entering into the Anderson-Holstein-like hamiltonian of the system \cite{Delbecq:11}, $g_{1(2)} \widehat{N}_{1(2)}(a+a^{\dag})$, $\widehat{N}_{1(2)}$ being the total number of electrons in QD1(2) and $a(a^{\dag})$ being the annihilation(creation) operators for the photons in the cavity at the fundamental frequency. This will allow us below to predict the expected form and magnitude of the interaction between the two dots which are dominated by single photon processes. Such a study is shown in figure 3a and 3b for QD1 and QD2 respectively. The colorscale plot of the differential conductance of the dots as a function the source-drain bias and the input power displays a characteristic funnel shape showing the conductance peak splitting in QD1 and QD2 as the power is increased. According to the picture developed previously the splitting should scale linearly with the amplitude of the field, or equivalently with the square root of the average number of photons. The latter viewpoint is particularly convenient since it allows us to determine directly the coupling constants of QD1 and QD2 to a single photon, $g_1$ and $g_2$ respectively. Indeed, the splitting $\Delta_{1(2)}$ reads : $\Delta_{1(2)}=2 g_{1(2)} \sqrt {\bar{n}}$, where $\bar{n}$ is the average number of photons in the cavity. The dependence of $\Delta_{1(2)}$ as a function of $\sqrt{\bar{n}}$ is shown in the inset of figure 3a and 3b for QD1 and 2 respectively. The linear behavior observed fully confirms our level driving picture and allows us to determine $g_{1} = 97 \textrm{MHz} \pm 22 \textrm{MHz}$ and $g_{2} = 98 \textrm{MHz} \pm 22 \textrm{MHz}$. We emphasize again that we obtain them by reexpressing the driving amplitude for each dot using $\lambda_{1(2)} V_\textrm{AC}=g_{1(2)} \sqrt {\bar{n}}$. The uncertainty in the determination of $g_{1(2)}$ is mainly related to the systematic errors made in the determination of the exact power at the input of our cavity.

\textbf{Gate tunable electron-photon coupling.}
We now investigate with greater detail the coupling mechanism between the photons and the electrons on the carbon nanotubes.  This section will allow us first to demonstrate interesting "toggles" for the electric control of the electron-photon coupling strength. Importantly, it will allow us to rule out all the spurious DC electrostatic effects which could hinder the study of the photon mediated distant interaction between the dots. We focus on QD1 for which electronic interactions can be neglected. The AC response of QD1 can be conveniently predicted with a self-consistent theory of AC transport similar to that of reference \cite{Buttiker:93}(see Methods). We measure $g_{1}$ for various gate voltages $V_\textrm{g1}$ using the previously described method. The result is presented in figure 4 b in inset by the filled dots. The value of $g_{1}$ displays modulations as $V_\textrm{g1}$ is swept. Strikingly, the change in $g_{1}$ is proportional to the linear conductance $G_1$ of QD1 as shown in figure 4b main panel. The value of $g_{1}$ increases by almost $100 \textrm{MHz}$ when the conductance $G_1$ is increased from $0.3 \times 2 \textrm{e}^2/\textrm{h}$ to $1.1 \times 2 \textrm{e}^2/\textrm{h}$. Such an increase is accounted for by the increase of the quantum capacitance of QD1, which is directly proportional to the conductance $G_1$ (see Methods). As shown in figure 4b, the sign of the slope of $g_1$ vs $G_1$ allows to discriminate between direct coupling between the cavity photons to the dot's energy level and indirect coupling to the dot's energy level via the source-drain electrodes. As expected from the efficient screening of the source-drain electrodes, the coupling mechanism is mainly via the source drain leads, as shown by the positive slope of $g_1$ vs $G_1$ (blue continuous line). One can further specify this by comparing the intercept of the blue line with the vertical axis. It corresponds to the geometrical part of the coupling strength of dot 1 since it corresponds to zero conductance. We find $48\textrm{MHz}$. An upper bound of the direct coupling can be obtained by reading off from figure 1d the DC side gate lever arm, of about $0.01$. This corresponds to only $2\textrm{MHz}$ and supports further that the coupling mechanism involves the source and drain leads and that figure 4a best displays the corresponding circuit diagram. Note that this mechanism is only possible at the frequency of the cavity since the source and drain electrodes are DC shunted in order to measure simultaneously the low frequency conductance (below 1kHz). Therefore, DC potentials on the central conductor of the cavity do not affect the energy levels of QD1 and any change in QD1's potential does not affect the potential of the central conductor.

\textbf{Distant coupling between the dots.}
We now probe the coupling between the states of QD1 and those of QD2, which is the central result of this article. We first use QD2 as a detector and measure the evolution of its energy levels via transport spectroscopy as its own gate and \textit{the distant gate of QD1} are swept. We present in figure 5a the differential conductance of QD2 as a function of $V_\textrm{g1}$ and $V_\textrm{g2}$ for one specific region. Strikingly, the gate of QD1 acts on QD2 even though the direct capacitive coupling is vanishingly small-the wide central conductor has a capacitance to the ground which is of about $0.7\textrm{pF}$, more than 5 orders of magnitude bigger than the gate capacitances of about $1\textrm{aF}$. Furthermore, as discussed above, the two dots, separated by the central conductor of the cavity, are only AC coupled to the cavity, via the source-drain leads. This strongly points to a distant interaction between the two dots without any direct electrostatic or tunnel coupling. In particular, it is not an electrostatic interaction mediated by a floating gate\cite{Trifunovic:12}.

Both quantum dots are coupled to the fundamental mode of the cavity. They can a priori be coupled via the virtual exchange of photons in a similar fashion as in atomic physics \cite{Raimond:01,Majer:07,Cottet:11}. The equivalent interaction mechanism, found in condensed matter whenever electronic states are coupled to bosonic modes, leads to a polaronic shift of each of the levels of QD1(2) : $\Delta^\textrm{polaron}_{1(2)}=-4 \pi g_1 g_2 N_{2(1)} / f_0 $, $N_{2(1)}$ being the number of electrons in QD2(1) respectively \cite{Mahan:78,Lehur:10}. This shift can be simply found from a Lang-Firsov transformation of the dot's total hamiltonian (see Methods). Since QD1 can contain a large number of electrons, $\Delta^\textrm{polaron}_{2}$ can become large, comparable to the energy scales relevant for QD2. This shift therefore survives at our moderately low temperatures even though the essence of this interaction is mainly quantum mechanical. One finds for example $\Delta^\textrm{polaron}_{2} \approx -1 \textrm{meV}$ for $N_1 \approx 10^4$. As shown in figure 4, $g_1$ and $g_2$ strongly depend on the gate voltage  of QD1 and QD2 respectively. This means that even though $N_{2(1)}$ are large numbers and do not vary substantially as the gates of each dot are swept, the shift $\Delta^\textrm{polaron}_{1(2)}$ can still evolve thanks to the gate tunability of the $g's$ as $V_\textrm{g1(2)}$ are varied. We now use $g_{1(2)} \approx g_{1(2)}^{0}-\alpha_{1(2)} V_{g1(2)}$ with the range of $\alpha_{1(2)}$ extracted from figure 4. Note that the hierarchy of energy scales here is the same as for phonons in solids which are usually much slower than the electrons, the electrons being delocalized over many sites. Physically, electrons on QD1 and QD2 shift the electric field corresponding to the cavity photons. The modification of the circuit energy by this effect is twofold. First, since QD2 is coupled to the cavity, the electrons of QD2 directly feel the photonic electric field deformation induced by the electrons of QD1. Second, since the energy of the cavity involves quadratically the photonic electric field, the shift of this electric
field by electrons of both QD1 and QD2 produces crossed terms between QD1 and QD2. These two phenomena lead to a global effective interaction between the two dots' electronic states.

In addition to the shift, the coupling of electrons to photons can lead to a splitting of the energy levels in each dot, proportional to the coupling constant of each dot to the photons. This can arise for instance due to photon induced K-K' mixing as predicted recently \cite{Palyi:11}. The general form of the energy levels of QD2 is therefore $\epsilon^{\pm}_{2} = \epsilon_0+ \alpha_0 V_\textrm{g2}-4 \pi g_1 (V_\textrm{g1}) g_2(V_\textrm{g2}) N_{1} / f_0 \pm g_2 (V_\textrm{g2}) \delta$, $\epsilon_0$ being a constant, $\alpha_0$ characterizing the bare coupling of QD2 to its gate and $\delta$ characterizing the magnitude of the splitting. The gate dependence of $g_2$ implies a variation of the splitting as gate 2 and gate 1 are swept, which explains the crossings of conductance peaks observed in figure 5a. As shown in figure 5b, we are able to reproduce with a simple modelling all the features observed in figure 5a (see Methods). The values of $\alpha_{1(2)}$ and $N_1$ are the three fitting parameters used to produce the main slope in figure 5b (see methods). The origin of the crossings is sketched schematically in figure 5c.

The distant coupling can be used to control in a non-local manner the electron-photon coupling strength, since it shifts the energy levels of each dot. We illustrate this fact by studying the evolution of $g_2$ as a function of $G_1$, as shown in figure 4e. The open circles are obtained for zero $V_\textrm{sd}$ and different gate voltages whereas all the other points are obtained by varying the source-drain bias at constant gate voltage. The systematic variation (roughly linear) of $g_2$ as a function of $G_1$ in the two different conditions depicted above shows that it is $G_1$ and therefore the quantum capacitance of dot 1 which matters for the shift of the energy levels rather than the gate voltage $V_\textrm{g1}$ itself. As a consequence, it further supports our distant coupling mechanism. As shown in figure 5a, we can follow the evolution of the energy levels of QD2 in the ($V_\textrm{g2}$-$V_\textrm{g1}$) plane. We present in figure 4c and d the bias spectroscopy of QD2 along one of the tilted lines of figure 5a for two different sets of ($V_\textrm{g2}$-$V_\textrm{g1}$). The size of the Coulomb diamonds is different for each of the sets of ($V_\textrm{g2}$-$V_\textrm{g1}$) which are conveniently expressed in terms of $G_1$ for similar reasons as above for the study of the local control of $g_1$. Measuring $g_2$ along this line gives the open symbols of figure 4e. For each of these points, we extract the capacitive lever-arm for the dot from the slope of the Coulomb diamonds. This allows us to predict the dependence of $g_2$, according to the coupling mechanism described in the Methods section. The resulting theoretical curve is the black solid line which accounts very well for our measurements of $g_2$.

\textbf{Discussion.}
Since both quantum dots are coupled to the microwave field, one can perform a joint read-out of their state by measuring the phase of the transmitted microwave signal through the cavity. Such a measurement is presented in figure 5d. The levels of QD1 and QD2 are clearly observed as tilted narrow and almost vertical shallow stripes respectively in the colorscale plot of the phase at $5.75 \textrm{GHz}$ as a function of $V_\textrm{g1}$ and $V_\textrm{g2}$. As expected, the levels of QD1 are wider than those of QD2. Their dependance on the distant gate (of QD2 now) is weaker than for QD2. We explain this fact by the smaller number of electrons $N_2$ in the closed quantum dot QD2. Note that the levels of QD1 and QD2 cross which means that they are not coherently coupled by the photons as expected because the "gamma's" are too large. The origin of the observed interaction relies however on the exchange of virtual photons like for the mechanism leading to two quantum bit operations for Rydberg atoms or superconducting quantum bits \cite{Raimond:01,Majer:07,Burkard:06}. Since the $g_1,g_2$ electron-photon couplings are the same as for the transmon-type quantum bits, we can therefore envision to perform SWAP operations as fast as for superconducting quantum bits. This will require converting the coupling to the charge to a coupling to the the spin \cite{Cottet:10,Jin:11} and using more isolated quantum dots.

\textbf{METHODS}\\

\textbf{Experimental}\\
Two single quantum dot circuits are fabricated using two \textit{different} and \textit{well separated} single wall carbon nanotubes connected to normal (Pd) electrodes, inside an Al microwave cavity shown in figure 1a (see ref \cite{Delbecq:11} for details). Both dots share the same anti-node of the electric field of the fundamental mode of the resonator, but they are located on opposite sides of the central conductor of the resonator, as shown in figure 1c. Their lateral size defined by nanolithography is about $400\textrm{nm}$, which is more than two orders of magnitude below the distance between them, about $80 \mu \textrm{m}$. The characteristics of the resonator are shown in figure 1b. The fundamental mode frequency is at $5.75 \textrm{GHz}$ with a quality factor of about $40$, as shown by the peak in the amplitude of the transmitted signal in red line. The phase of the microwave signal, shown in black line, displays the characteristic jump of $\pi$ at the resonance. The temperature of the experiment is $1.5\textrm{K}$ throughout the paper.

\textbf{Modelling of the evolution of the conductance peaks.}\\
One can model the influence of the level structure depicted in the main text on transport in a Coulomb blockaded quantum dot as QD2. We present in figure 5b the result of such a modeling using a simplified equation of motion theory for the Green's functions and the Meir-Wingreen formula for the linear conductance \cite{Meir:92}. The colorscale plot of figure 5b has been obtained using using $N_1=10^6$ and $\alpha_{1(2)} \approx 25 \textrm{MHz/V}$ in the $\{V_\textrm{g1},V_\textrm{g2}\}$ region considered here. In particular, the alternation between level crossings and splittings as gate 2 is swept towards negative values is reproduced. This phenomenon can be explained qualitatively by the picture of figure 5c. In a quantum dot in the Coulomb blockade regime, a splitting of energy levels in the intrinsic spectrum leads to a separation between conductance peaks corresponding to the same orbital and to crossing of conductance peaks corresponding to different orbital levels.

\noindent\textbf{Electron-photon coupling mechanism}\\
The electrostatics of the dots allows us to determine the level energy at the mean-field level \cite{Buttiker:93}: $\xi_\textrm{d} = \epsilon_\textrm{d} - e \alpha_\textrm{AC} V_\textrm{AC} - e (\alpha_\textrm{L}+\alpha_\textrm{R})\tilde{\alpha}_\textrm{AC} V_\textrm{AC}+ E_\textrm{C} \langle N \rangle$ where $\langle N \rangle$ is the average number of charge on the island, $\epsilon_\textrm{d}$ is the bare dot energy and $E_\textrm{C}=\frac{e^2}{2C_\Sigma}$ is the charging energy. The other important terms in the above expression are : $\alpha_\textrm{AC} $ the direct cavity-dot capacitive coupling, $\alpha_\textrm{L,R}$ the $L,R$ capacitances of the source-drain electrodes and $\tilde{\alpha}_\textrm{AC}$ is the source(drain) lead -cavity capacitive coupling which we assume to be the same for source and drain here (symmetric coupling). The self-consistent equation that governs the number of charges on the dot is $\langle N \rangle = \int d\epsilon f(\epsilon - \mu_\textrm{ec}) \mathcal{A}(\epsilon - \xi_\textrm{d})$, with $f$ the Fermi function, $\mu_\textrm{ec}$ the electrochemical potential of the leads and $\mathcal{A}$ the spectral density of the dot. The electro-chemical potential of the dot is changed via the AC drive $V_\textrm{AC}$ of the cavity as $\mu_\textrm{ec} = \mu -e \tilde{\alpha}_\textrm{AC} V_\textrm{AC}$. From the above equations, one can derive $\frac{\partial \langle N \rangle}{\partial V_\textrm{AC}} =\{ e \alpha_\textrm{AC} + e (\alpha_\textrm{L}+\alpha_\textrm{R}-1)\tilde{\alpha}_\textrm{AC}- E_\textrm{C} \frac{\partial \langle N \rangle}{\partial V_\textrm{AC}}\}\times -\frac{\partial \langle N \rangle}{\partial \epsilon_\textrm{d}}$. The electron-photon coupling then is obtained by the derivative of the dot energy level with respect to the fluctuating potential of the resonator :
\begin{align}
g= V_\textrm{rms} \frac{\partial \xi_\textrm{d}}{\partial V_\textrm{AC}} = e V_\textrm{rms} \{\alpha_\textrm{AC}+(\alpha_\textrm{L}+\alpha_\textrm{R})\tilde{\alpha}_\textrm{AC} - E_\textrm{C} \times -\frac{\partial \langle N \rangle}{\partial \epsilon_\textrm{d}} \frac{\alpha_\textrm{AC}+(\alpha_\textrm{L}+\alpha_\textrm{R}-1)\tilde{\alpha}} {1 + E_\textrm{C} \times -\frac{\partial \langle N \rangle}{\partial \epsilon_\textrm{d}}}\}
\end{align}

The charge susceptibility $\frac{\partial \langle N \rangle}{\partial \epsilon_d}$ of the dot is the opposite of the quantum capacitance $C_Q$ of the dot in the non-interacting limit ($\frac{\partial \langle N \rangle}{\partial \epsilon_\textrm{d}}=-C_\textrm{Q}/e^2$) and is proportional to the linear conductance $G$ of the dot \cite{Buttiker:93}. If the dot is directly coupled to the cavity, the above expression shows a negative slope as a function of $G$ whereas if the dot is indirectly coupled via the leads, the above expression shows a positive slope which arises from the sum rule $\alpha_\textrm{L}+\alpha_\textrm{R}<1$. This allows us to plot the blue line of figure 4b and to conclude that our dots are coupled to the cavity photons via the dot's source-drain leads.\\

\noindent\textbf{Distant coupling and Lang-Firsov transformation}\\
We reproduce here the Lang-Firsov transformation which allows us to derive the interaction between the distant dots. We first consider the Anderson-Holstein hamiltonian of the main text :
\begin{align}
H  &  =H_{0}+H_\textrm{c}\\
H_{0}  &  =\hbar\omega_{0}a^{\dag}a+\sum\limits_{d\sigma}\varepsilon_{d\sigma
}c_{d\sigma}^{\dag}c_{d\sigma}+H_{int}\\
H_{c}  &  =\sum\limits_{d\sigma}g_{d}(a^{\dag}+a)c_{d\sigma}^{\dag
}c_{d\sigma}%
\end{align}

$d\ $\ corresponds to the different dots. We assume $g_{d}$ dependent on the dot $d$ since the dots are different and located in different positions. The operator $H_{int}$ accounts for the electronic interaction on each dot.

We follow ref \cite{Mahan:78} and use the unitary transformation for an operator $A$ :

\begin{equation}
\widetilde{A}=e^{S}Ae^{-S} \label{Htild}%
\end{equation}
with
\begin{equation}
S=\sum_{d\sigma}\frac{g_{d}}{\hbar\omega_{0}}(a^{\dag}-a)c_{d\sigma
}^{\dag}c_{d\sigma}%
\end{equation}

Using the identity :
\begin{equation}
\widetilde{O}=O+[S,O]+\frac{1}{2!}[S,[S,O]]+\frac{1}{3!}[S,[S,[S,O]]]+....
\label{develop}%
\end{equation}

we get :

\begin{equation}
\widetilde{H}  =\hbar\omega_{0}a^{\dag}a-\frac{1}{\hbar\omega_{0}}\left(  \sum_{d\sigma
}g_{d}n_{d\sigma}\right)  ^{2}+\sum\limits_{\sigma}\varepsilon_{d\sigma
}n_{d\sigma}+H_{int}%
\end{equation}\\

The second term contains two quadratic terms and one bilinear term in the number of electrons occupying each dot. While the first can be absorbed in the interaction term of hamiltonian of each dot, the bilinear term can be factorized in each orbital part of the hamiltonian, leading to the expression used in the main text for the polaronic shift. This leads to the distant coupling of the main text. Note that, contrarily to transport where only the states close to Fermi energy matter, it is the total charge which is coupled to the electromagnetic field (the sum goes over all occupied states). This can lead to a very large shift of the levels of each dot provided the distant dot has a high enough charge density. In addition, contrarily to its atomic physics counterpart, the above expression is non-perturbative and is exact when the dot-lead coupling is zero.

Using the above equation, we get :

\begin{equation}
\epsilon_{2} = \epsilon_0+ \alpha_0 V_\textrm{g2}-4 \pi g_2(V_\textrm{g2}) \sum_{d\sigma
}g _{1,j}n_{1,j\sigma} / f_0 %
\end{equation}\\

This can be further simplified in the continuum limit as :

\begin{equation}
\epsilon_{2} = \epsilon_0+ \alpha_0 V_\textrm{g2}-\frac{4 \pi g_2(V_\textrm{g2})}{f_0} N_1 \frac{\int^{V_\textrm{g1}} d \epsilon g_1(\epsilon) \textsl{A}_\textrm{QD1}(\epsilon)}{\int^{V_\textrm{g1}} d \epsilon \textsl{A}_\textrm{QD1}(\epsilon)}%
\end{equation}\\

where $N_1= \int^{V_\textrm{g1}} d \epsilon \textsl{A}_\textrm{QD1}(\epsilon)$ for QD1. By defining $ \langle g_1 \rangle=\frac{\int^{V_\textrm{g1}} d \epsilon g_1(\epsilon) \textsl{A}_\textrm{QD1}(\epsilon)}{\int^{V_\textrm{g1}} d \epsilon \textsl{A}_\textrm{QD1}(\epsilon)}$, the implicit equation for $\{V_\textrm{g2},V_\textrm{g1}\}$ reads :
\begin{equation}
\epsilon_0+ \alpha_0 V_\textrm{g2}-\frac{4 \pi g_2(V_\textrm{g2})\langle g_1 \rangle (V_\textrm{g1})}{f_0} N_1  = const
\end{equation}

This is the most important part of the equation of the main text and controls the dispersion of the energy levels of dot 2. Since $\langle g_1 \rangle (V_\textrm{g1})$ is an integral over all the occupied energy levels, its variations are smooth and roughly linear over a large gate voltage scale. The variations of $g_2(V_\textrm{g2})$ can be more pronounced and can lead in general to a non-linear dispersion of the energy levels in the $\{V_\textrm{g2},V_\textrm{g1}\}$ plane, as observed in the joint read-out of figure 5d.

Using the values given in the maintext, we are able to reproduce the observed general slope of the energy levels of QD2 in the $\{V_\textrm{g2},V_\textrm{g1}\}$ plane. Note that the above analysis, carried out for a single mode of the electromagnetic field, can be straightforwardly extended to the multi-mode case. Since the coupling mechanism demonstrated here is non-resonant, essentially all the modes within the line-width of the energy levels can contribute\cite{Lehur:10}. This can reduce the number of electrons needed to explain our data by a factor of 100-300.

{\noindent\small{\bf Acknowledgements. } We are indebted to P. Simon, C. Bergenfeldt, P. Samuelsson C. Mora, K. Le Hur and G. Zarand for fruitful discussions. The devices have been made within the consortium Salle Blanche Paris Centre. This work is supported by the ANR contracts DOCFLUC, HYFONT, SPINLOC and the EU-FP7 project SE2ND[271554].

{\noindent\small{\textbf{Competing financial interests.}} The authors declare no competing financial interests.

\noindent\textbf{Authors contributions}\\
M.R.D. and T.K. conceived the experiment, carried out the measurements and analyzed the data with inputs from A.C. and J.J.V.; M.R.D. carried out the nanofabrication process with inputs from L.E.B. and S.D.; A.C. provided theoretical material ; M.R.D. and T.K. co-wrote the manuscript with the help of A.C.\\

\begin{figure}[!pth]
\centering\includegraphics[height=0.85\linewidth,angle=0]{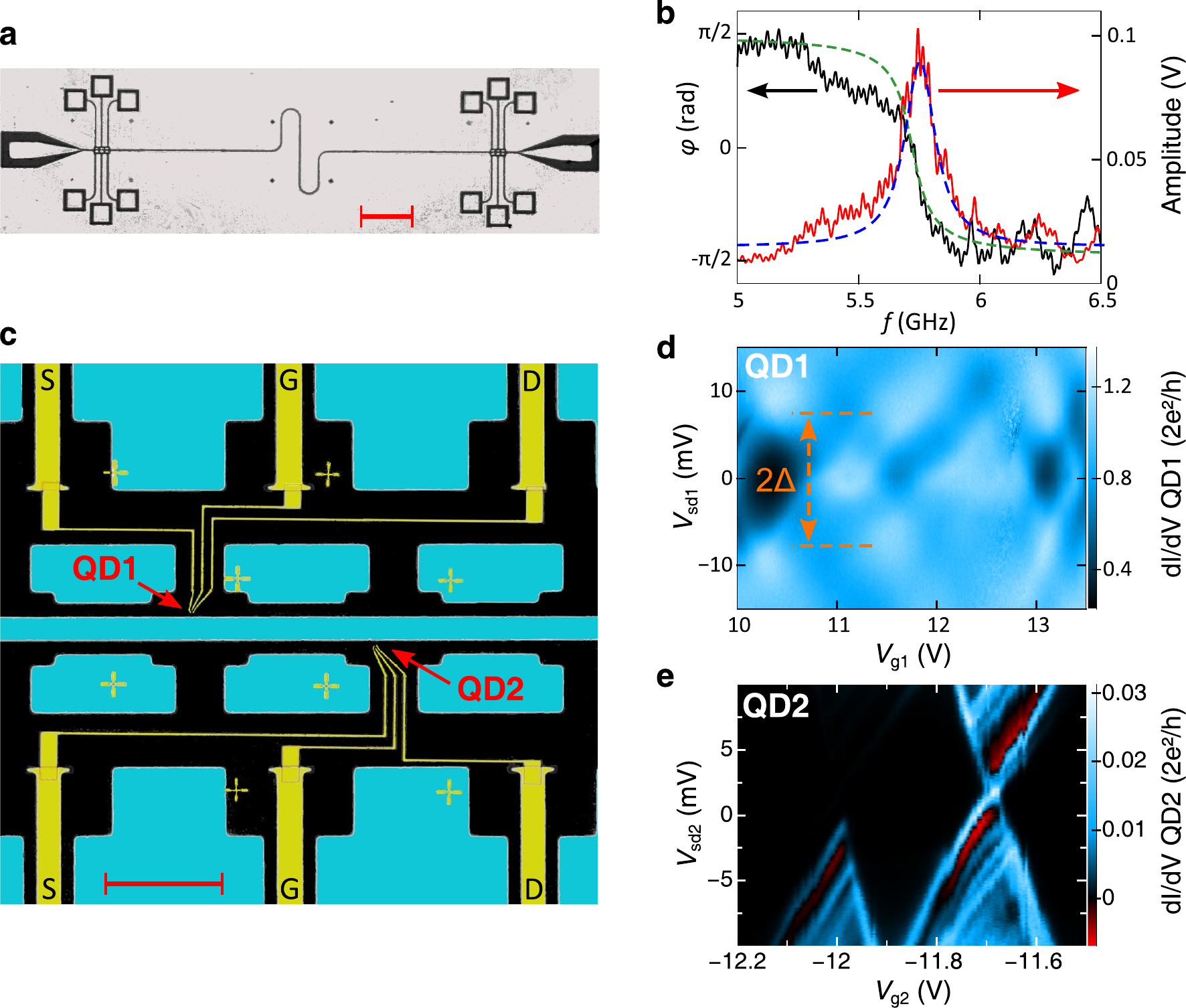}
\caption{\textbf{The coupling architecture.} \textbf{(a)} Optical micrograph of the Al microwave resonator in coplanar waveguide geometry. 4 sets of 3 DC lines allow to contact up to 4 QDs inside the cavity. The bar is 1mm. \textbf{(b)} Transmitted amplitude (red curve) and phase (black curve) of the microwave field versus frequency, giving $f_0=5.75$GHz and  $Q\approx 40$. \textbf{(c)} False colors optical micrograph showing the two quantum dots structures inside the microwave cavity. Aluminium elements of the resonator are represented in blue (transmission line and ground plane). The Pd DC lines contacting the carbon nanotubes (not visible here), are represented in yellow. The two QDs are  separated by 80$\mu$m and are designed by the standard source-drain and side gate geometry. The bar is 50 $\mu \textrm{m}$. \textbf{(d)} Color scale plot of QD1 differential conductance in units of $2\textrm{e}^2/\textrm{h}$ in the gate voltage $V_\textrm{g1}$ bias voltage $V_\textrm{sd1}$ plane. The spectroscopy shows a Fabry-Perot behavior with a level spacing of 7 meV, corresponding to a 200 nm long quantum dot. \textbf{(e)} Color scale plot of QD2 differential conductance in units of $2\textrm{e}^2/\textrm{h}$ in the gate voltage $V_\textrm{g2}$-bias voltage $V_\textrm{sd2}$ plane. The spectroscopy shows a Coulomb diamond with excited states.
}%
\label{device}
\end{figure}

\begin{figure}[!pth]
\centering\includegraphics[width=1\textwidth]{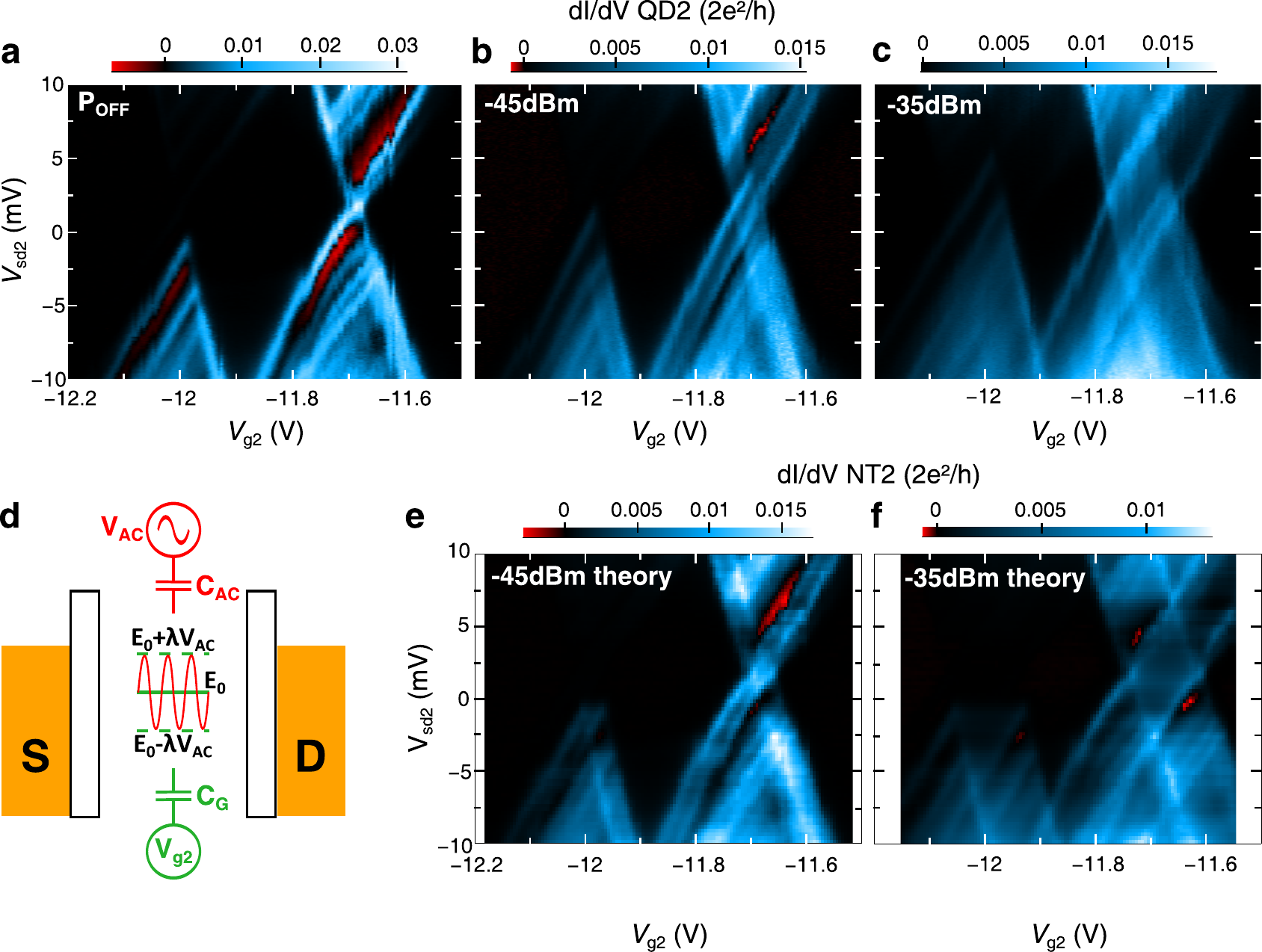}
\caption{\textbf{Driving quantum dot 2 through the cavity.} Influence of the microwave power driving the cavity on QD2 spectroscopy. Differential conductance of QD2 as a function of the gate voltage $V_\textrm{g2}$ and the bias voltage $V_\textrm{sd2}$ for different powers at the input port of the cavity : \textbf{(a)} no power, \textbf{(b)} $P_\textrm{in}$=-45dBm and \textbf{(c)} $P_\textrm{in}$=-35dBm. The conductance peaks of the quantum dot split, as emphasized by the formation of a new diamond shaped region around $V_\textrm{g2}=-11.7$V. \textbf{(d)} Scheme showing the capacitive coupling of the quantum dot electronic levels to the gate electrode (green) and to the resonator (red). The DC gate electrode drives the level energy $E_0$ in the quantum dot. The AC field of the resonator modulates the energy levels with an amplitude $\lambda V_\textrm{AC}$ and frequency $f_0$. The averaging of this modulation leads to two split conductance peaks (dashed lines) at $E_0 \pm \lambda V_\textrm{AC}$. Simulations of the splitting based on the scheme depicted in \textbf{(d)} and applied to the spectroscopy of \textbf{(a)} with $\lambda=15.5$, respectively for $P_\textrm{in}$=-45dBm \textbf{(e)} and $P_\textrm{in}$=-35dBm \textbf{(f)}.
}%
\label{greyscale}%
\end{figure}

\begin{figure}[!pth]
\centering\includegraphics[height=0.65\linewidth,angle=0]{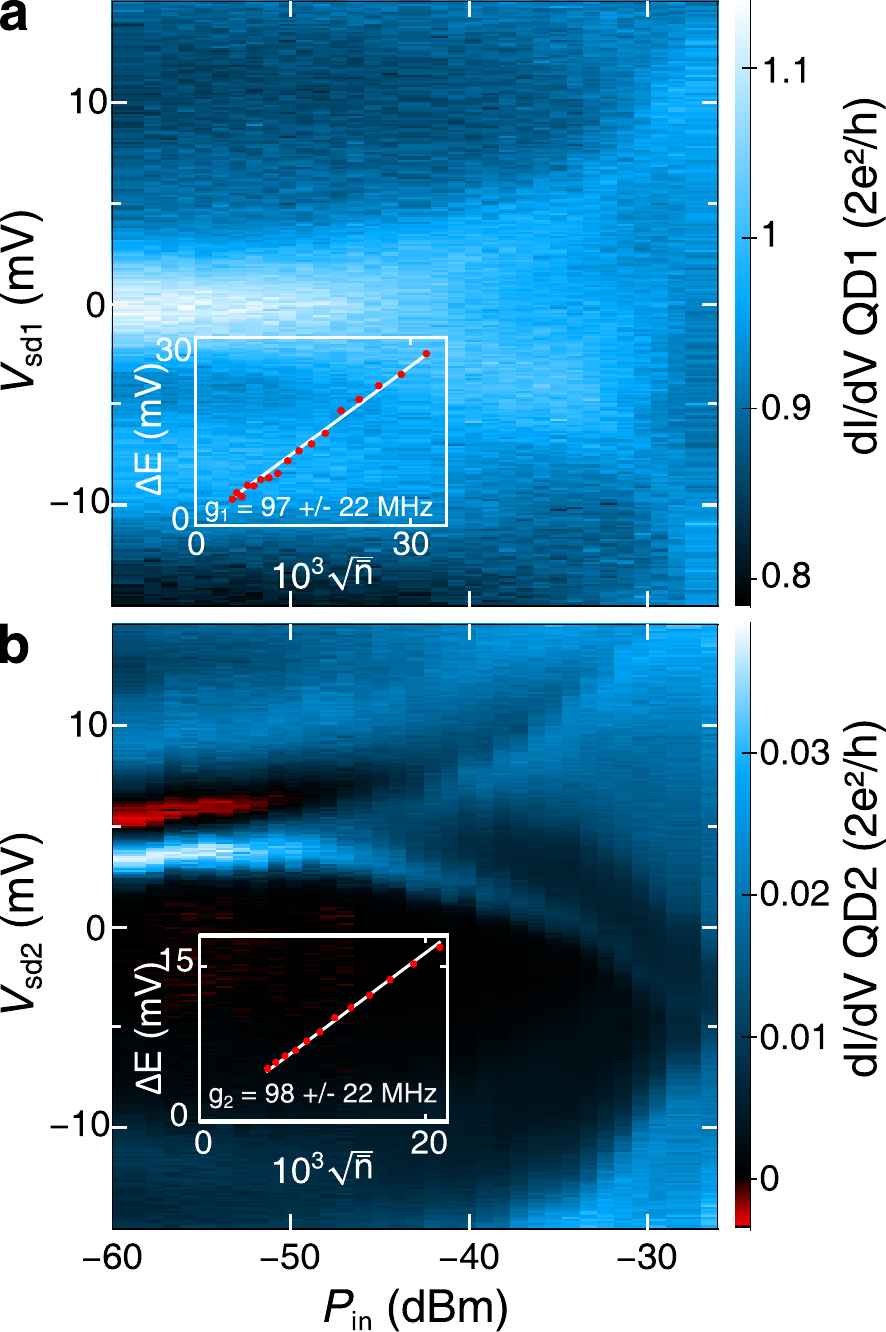}
\caption{\textbf{Measurement of the electron-photon coupling strength for each dot.} Microwave power dependence of conductance peak splitting. \textbf{(a)} and \textbf{(b)} Color scale plots of the differential conductance of QD1 (respectively QD2) in the input power $P_\textrm{in}$-bias voltage $V_\textrm{sd1}$ (respectively $V_\textrm{sd2}$) plane. Insets : the dependence of the conductance peak separation $\Delta E$ versus $\sqrt{\bar{n}}$ for each QD (red dots). The slope of the linear fit (white line) gives the coupling $g_{1(2)}$ between the quantum dot 1(2) and the microwave field as $\Delta E/h = 2g_{1(2)}\sqrt{\bar{n}}$.
}
\label{diagram}
\end{figure}

\begin{figure}[!pth]
\centering\includegraphics[height=0.65\linewidth,angle=0]{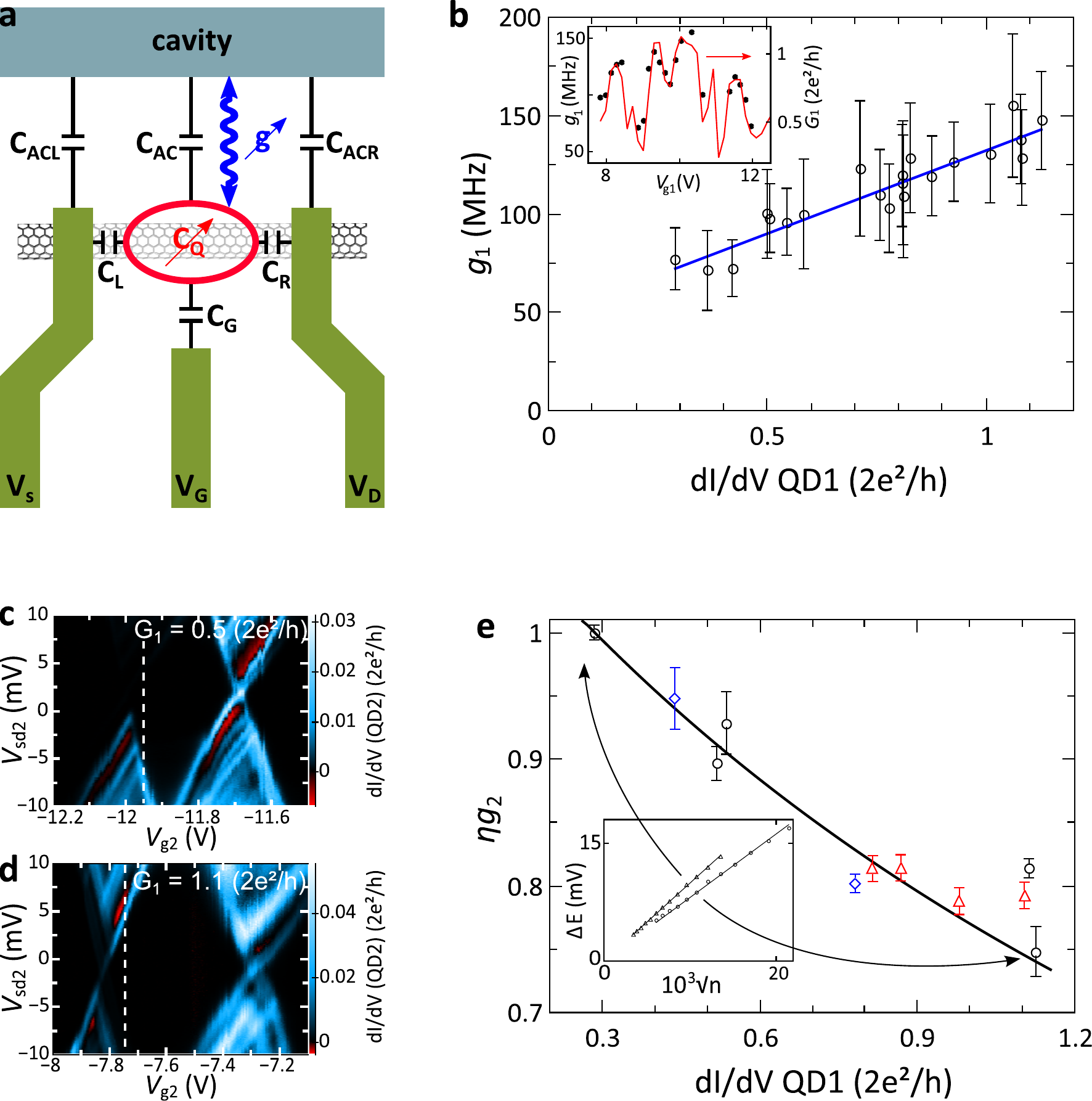}
\caption{\textbf{Coupling mechanism and tunable coupling strength} \textbf{(a)} Schematics of the coupling mechanism between the quantum dots and the cavity. \textbf{(b)} Variation of the electron-photon coupling strength of QD1, in open dots, as a function of its own linear conductance. The error bars correspond to the systematic error of the linear fitting for the plots in inset of figure 3. The solid line is the theory described in the methods section. Inset : Variations of the electron-photon coupling strength and of the conductance of QD1 as a function of $V_\textrm{g1}$. \textbf{(c)} and \textbf{(d)} Transport spectroscopy of QD2 for two different values of the conductance of QD1 (corresponding to two different sets of ($V_\textrm{g2}$,$V_\textrm{g1}$) for the same energy level of QD2. \textbf{(e)} Relative variation of the electron-photon coupling strength of QD2 as a function of the linear conductance of QD1. The open circles correspond to a change of the conductance of dot 1 by changing its gate. The open triangles and diamonds correspond to a change of conductance by changing the source drain bias. The solid black line is the theory. Inset : the dependence of the conductance peak separation $\Delta E$ versus $\sqrt{\bar{n}}$ for the two points indicated by the arrows. The error bars correspond to the systematic error of the linear fitting for the plots in inset of figure 3.
}
\label{tunability}
\end{figure}

\begin{figure}[!pth]
\centering\includegraphics[height=0.65\linewidth,angle=0]{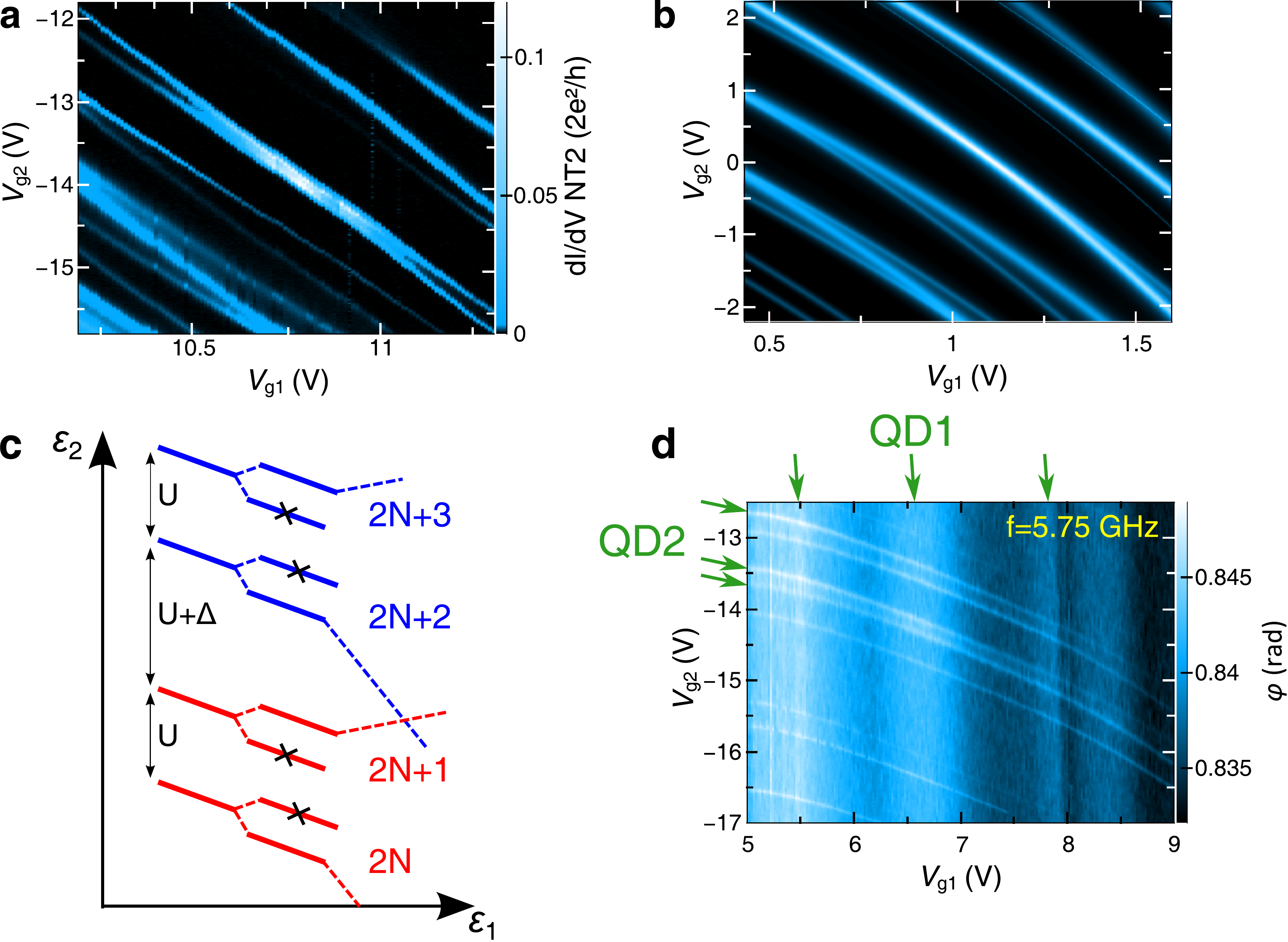}
\caption{\textbf{Distant coupling between the quantum dots.} \textbf{(a)} Color scale plot of the differential conductance of QD2 versus its own gate voltage $V_\textrm{g2}$ and the gate voltage $V_\textrm{g1}$ of QD1. The levels of QD2 evolve with different slopes, leading to level crossings. \textbf{(b)}. Modeling of the conductance map of QD2 using the form of  $\epsilon^{\pm}_{2}$ presented in the main text and a simplified equation of motion theory. \textbf{(c)} Diagram of the transport spectroscopy of QD2 versus $\epsilon_2$ and $\epsilon_1$, the energies of QD2 and QD1 respectively. Two orbitals are represented (red and blue levels), separated by $U+\Delta$, with $\Delta$ the orbital spacing. The two conductance peaks corresponding to the same orbital are separated by the charging energy $U$. The interaction between QD1 and QD2 shifts the levels of QD2. They are also split as shown by the dashed lines. The conductance peaks separated by less than $U$ in the same orbital are forbidden (shown by a cross). Crossings between conductance peaks originating from different orbitals are allowed. The resulting spectroscopic pattern exhibits an alternation of conductance peaks moving away and coming closer, as observed in \textbf{(a)}. \textbf{(d)} Joint readout of the 2 QDs in the phase $\phi$ of the transmitted RF signal. Both levels of both QD1 and QD2, indicated by green arrows, can be observed and cross.
}
\label{coupling}%
\end{figure}

\end{document}